\documentclass[aps,prl,reprint,superscriptaddress]{revtex4-1}

\usepackage[dvips]{graphicx} % for figures
\usepackage{amsfonts}
\usepackage{amssymb}
\usepackage{amscd}
\usepackage{amsmath,braket,bm}    % need for subequations
\usepackage{enumerate}
\usepackage{epsfig}
\usepackage{subfigure}
\usepackage{xcolor}
\usepackage{amsthm}
\usepackage{framed}
\usepackage{soul}

\usepackage{color}
\usepackage{hyperref}
\hypersetup{  colorlinks=true, linkcolor=blue, citecolor=red, urlcolor=blue  }

\usepackage{physics}

\definecolor{blue}{rgb}{0,0,1}
\definecolor{red}{rgb}{1,0,0}

\begin{document}
\title{Simulating Molecular Spectroscopy with Circuit Quantum Electrodynamics}
%\title{Simulating Molecular Spectroscopy with 3D Superconducting Circuit QED}
%\title{Simulating Molecular Dynamics with Superconducting Cavity QED}
%\title{Quantum Molecular Simulation with Superconducting Cavity QED}
%\title{Experimental Simulation of Molecular Transitions with Superconducting Cavity QED}
%\title{Simulating molecules in Superconducting Cavity QED}
%\title{Molecular Simulation in Superconducting Cavity QED}
%\title{Quantum simulation of molecular absorption spectra and its realization by circuit superconducting qubits}

\author{L. H$^*$}
\author{Y. C. Ma$^*$}
\author{Y. Xu}
\author{W. Wang}
\author{Y. Ma}
\author{K. Liu}
\affiliation{Center for Quantum Information, Institute for Interdisciplinary Information Sciences, Tsinghua University, Beijing 100084, China}
\author{M.-H. Yung$^\dagger$}
\email{yung@sustc.edu.cn}
\affiliation{Institute for Quantum Science and Engineering and Department of Physics, South University of Science and Technology of China, Shenzhen 518055, China}
\author{L.~Sun$^\dagger$}
\email{luyansun@tsinghua.edu.cn}
\affiliation{Center for Quantum Information, Institute for Interdisciplinary Information Sciences, Tsinghua University, Beijing 100084, China}

% It is always \today, today,
             %  but any date may be explicitly specified

\begin{abstract}

% One or two sentences providing a basic introduction to the field, comprehensible to a scientist in any discipline.
Spectroscopy is a crucial laboratory technique for understanding quantum systems through their interactions with electromagnetic radiation. 
% Two to three sentences of more detailed background, comprehensible to scientists in related disciplines.
Particularly, spectroscopy is capable of revealing the physical structure of molecules, leading to the development of the maser---the forerunner of the laser.
% One sentence clearly stating the general problem being addressed by this particular study.
However, real-world applications of molecular spectroscopy~\cite{kroto1992molecular} are mostly confined to equilibrium states, due to computational and technological constraints; a potential breakthrough can be achieved by utilizing the emerging technology of quantum simulation.  
% One sentence summarizing the main result (with the words ¡°here we show¡± or their equivalent).
Here we experimentally demonstrate that a superconducting quantum simulator~\cite{Houck2012a} is capable of generating molecular spectra for both equilibrium and non-equilibrium states, reliably producing the vibronic structure of the molecules.
% Two or three sentences explaining what the main result reveals in direct comparison to what was thought to be the case previously, or how the main result adds to previous knowledge.
Furthermore, our quantum simulator is applicable not only to molecules with a wide range of electronic-vibronic coupling strength characterized by the Huang-Rhys parameter~\cite{mukamel1999principles}, but also to molecular spectra not readily accessible under normal laboratory conditions. 
% One or two sentences to put the results into a more general context. 
These results point to a new direction for predicting and understanding molecular spectroscopy, exploiting the power of quantum simulation. 
 
\end{abstract}

\maketitle

Quantum simulation represents a powerful and promising means to overcome the bottleneck for simulating quantum systems with classical computers, as advocated by Feynman~\cite{Feynman1982}. One of the major applications for quantum simulation is to solve molecular problems~\cite{Aspuru-Guzik2005c,Kassal2011,Whitfield2011,CodyJones2012,Babbush2014}. In recent years, much experimental progress has been achieved in simulating the electronic structures of molecules using quantum devices. Particularly, the potential energy surface of the hydrogen molecule was simulated experimentally~\cite{Lanyon2010, Du2010, Wang2015}.
%using quantum optics~\cite{Lanyon2010} and nuclear magnetic resonance~\cite{Du2010}. Furthermore, the electronic structure for the helium hydride cation was simulated with a nitrogen-vacancy defect in diamond~\cite{Wang2015}. 
However, it remains a challenge to scale up this type of experiments for larger molecules, as the phase-estimation method involved requires an enormous amount of computing resources for implementation. 

An alternative and potentially more economical approach for quantum molecular simulation has been achieved by using a quantum variational approach~\cite{Yung2014,Peruzzo2013,Shen2015,o2015scalable} that aims to improve the eigenstate approximation through local measurements of the Hamiltonians. 
%This approach was experimentally demonstrated with on-chip quantum optics~\cite{Peruzzo2013}, trapped ions~\cite{Shen2015}, and superconducting qubits~\cite{o2015scalable}. 
So far, most (if not all) of the molecular simulation experiments performed are all confined to the study of static properties of molecules. It is still an experimental challenge to utilize quantum simulators for studying molecular dynamics, in particular, molecular spectroscopy. Furthermore, classical methods in predicting vibrationally-resolved absorption spectra are mostly limited in the gas phase. However, most chemical processes occur in solution, where the molecular vibrational motion depends heavily on the environment; predicting molecular spectroscopy for non-equilibrium states represents a major challenge in quantum chemistry~\cite{Improta2007}. 

\begin{figure*}
\includegraphics[scale=1]{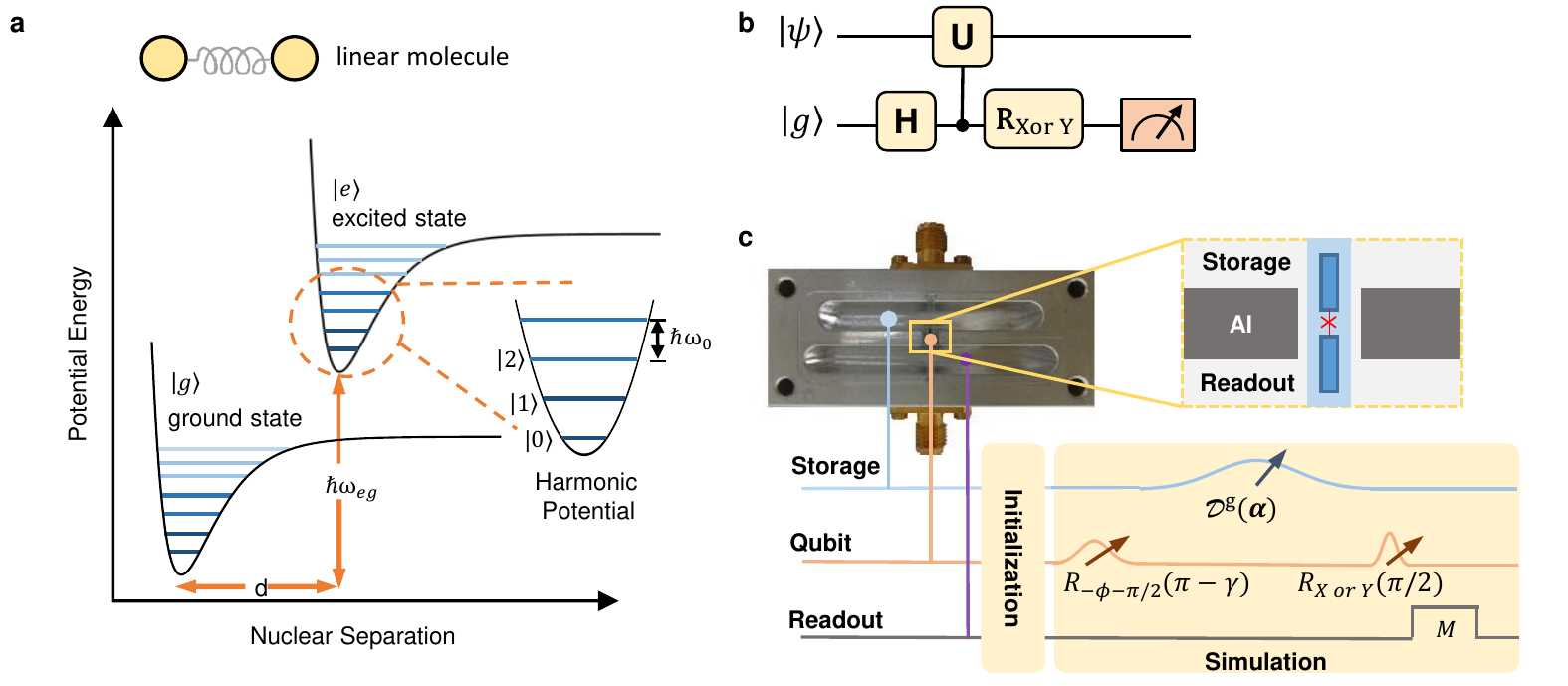}
\caption{\label{fig:double_level} \textbf{Basic principle of the superconducting simulator.} \textbf{(a)} {Two identical energy surfaces of a molecule, with one curve displaced from the other along a nuclear coordinate. Near the minimum, the energy surfaces  can be approximated by a harmonic potential, with an energy separation of $\hbar\omega_0$. Here $\hbar\omega_{eg}$ is the 0-0 energy splitting; in most cases $\omega_{eg} \gg \omega_0$.} \textbf{(b)} The kernel quantum circuit diagram of our method. The circuit consists of an ancilla qubit and a bosonic system. The bosonic system represents nuclear motion mode and is in an initial state $\ket{\psi}$. Similar to DQC1~\cite{knill1998power}, a composite {evolution gate} $U\equiv e^{iH_gt/\hbar}e^{-iH_et/\hbar}$ is applied to the system following a Hadamard gate on the qubit. Finally, measurements along $X$ and $Y$ axis are performed to obtain the correlation function $C_{\mu\mu}=\left\langle {{\sigma _x}} \right\rangle  + i\left\langle {{\sigma _y}} \right\rangle$. \textbf{(c)} Device layout and pulse sequence for the superconducting simulator. A ``vertical" transmon qubit (dark blue in the enlarged device schematic) on a sapphire chip (light blue) in a waveguide trench couples to two 3D Al cavities. The qubit is first prepared in the ground state $\ket{g}$ and the storage cavity (the bosonic system in \textbf{b}) is initialized to different states for various simulations (see the main text for details). As shown in \textbf{b}, the simulation scheme consists of three processes: a qubit rotation $R_{-\phi-\pi/2}(\pi-\gamma)$, a controlled displacement ${\cal D}^g(\alpha)$ of the cavity conditional upon the qubit state $\ket{g}$, and finally a $\sigma_x$ or $\sigma_y$ measurement of the qubit. {Here $R_{\varphi}(\theta)$ represents a $\theta-$rotation along $\varphi-$axis in $X$-$Y$ plane on Bloch sphere.}  Note that the rotation angle $\pi-\gamma$ in our simulation is not limited to $\pi/2$ (see the main text).}
\end{figure*}

\begin{figure*}
    \includegraphics[scale=1]{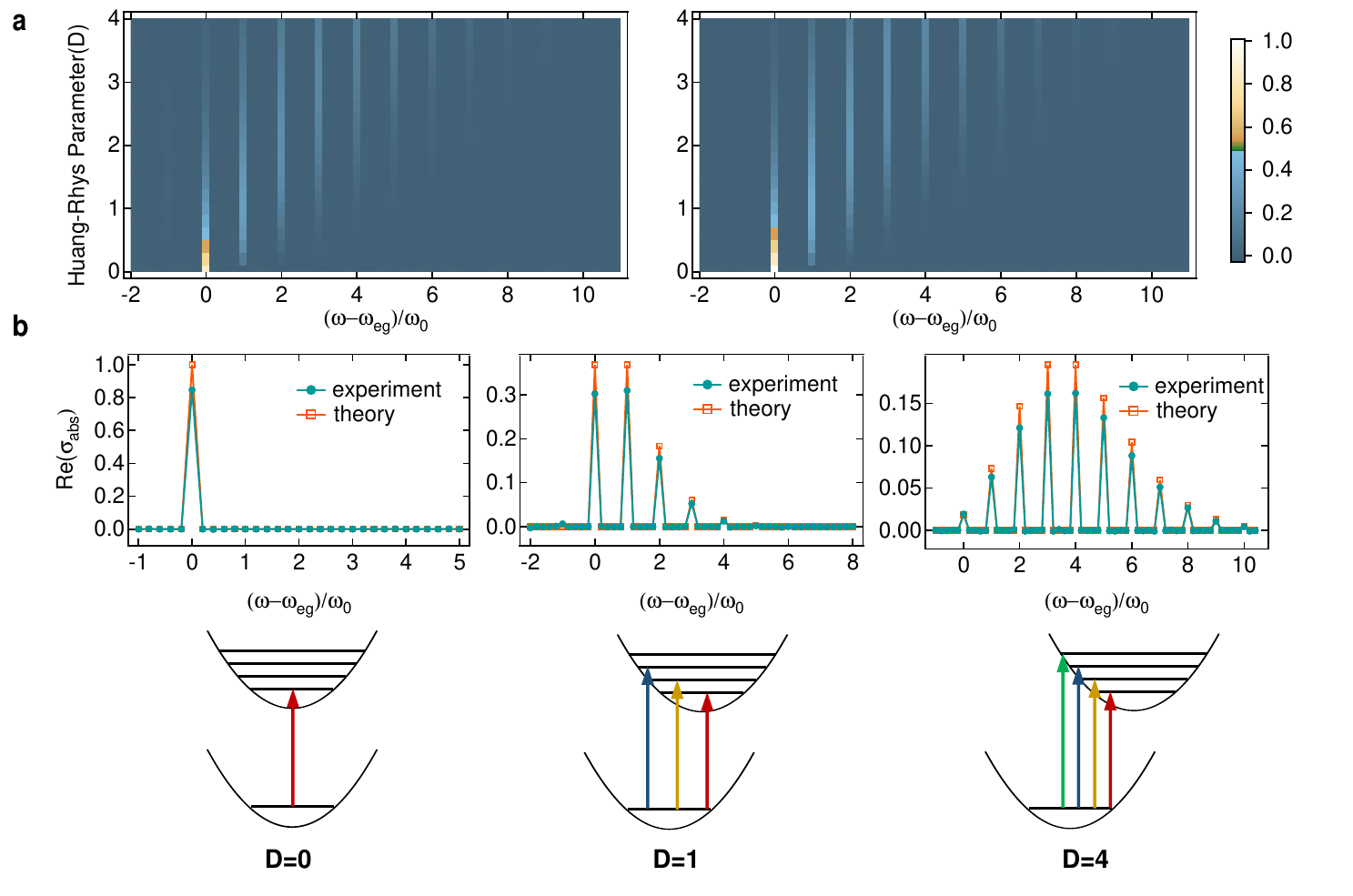}
  \caption{\label{fig:zero_temperature} \textbf{Absorption spectrum of the nuclear system at vacuum ($\ket{\psi}=\ket{0}$) at zero temperature}. \textbf{(a)} Progress of the absorption spectrum as a function of Huang-Rhys parameter $D$. Here we only present the real part of $\sigma_{abs}$. The $x$ axis represents the normalized spectral frequency of electronic transition, which describes the necessary energy for transitions from the electronic ground state to the excited states. We compare experimental results (left) with theory (right) using the same color scale which represents the transition probability. \textbf{(b)} The cross section at $D$=0, 1, and 4, and the corresponding schematics of the electronic transitions (bottom row). As expected, the absorption peaks of zero-temperature molecular spectrum arises from $\omega_{eg}$, separated by $\omega_0$ with a Poisson distribution of intensities. The experimental data are lower by a constant reduction factor $f=0.83$ than theory, as expected dominantly due to the decoherence of the qubit during the simulation process.}
\end{figure*}

\begin{figure*}
    \includegraphics[scale=1]{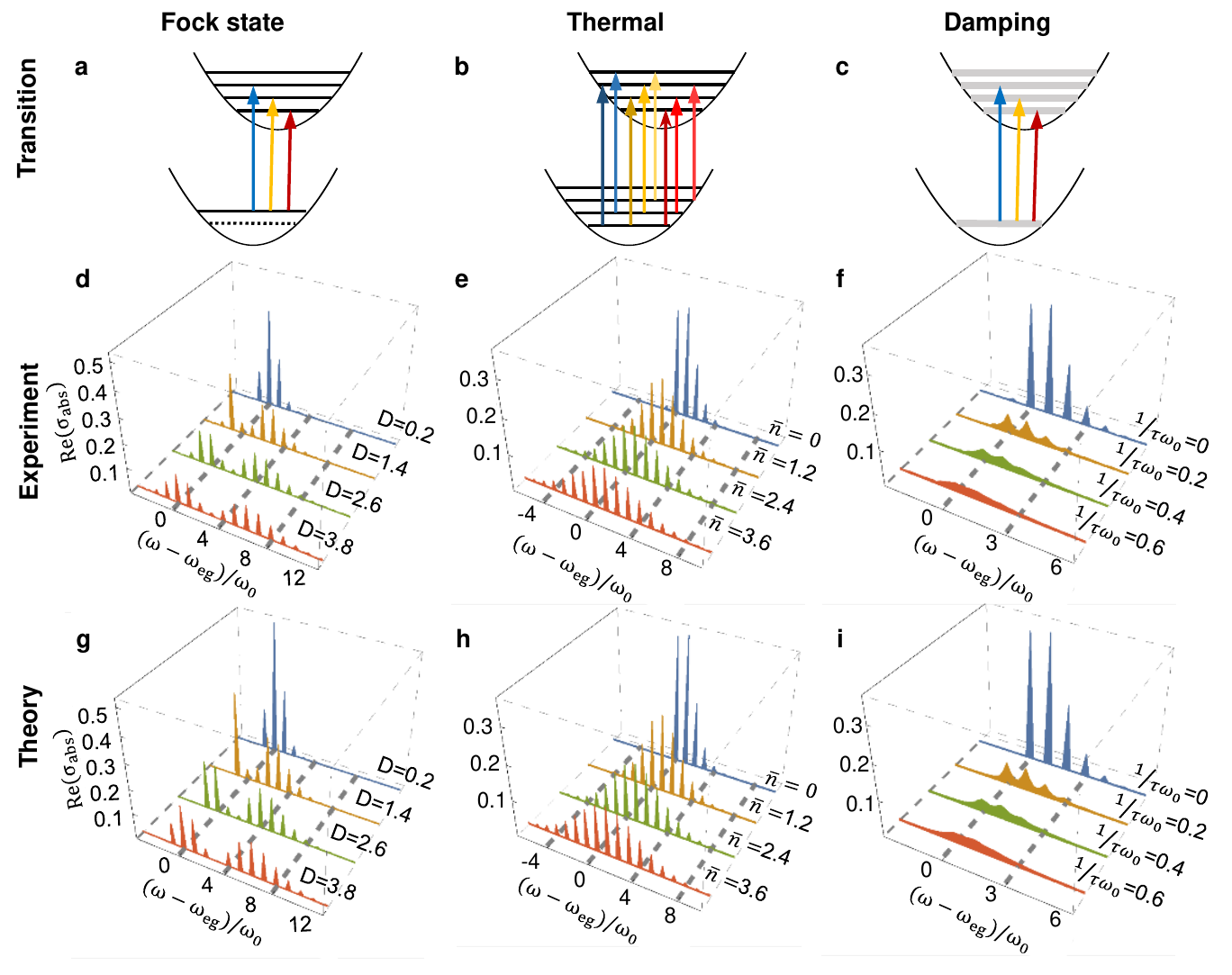}
    \caption{\label{fig:other_cases}\textbf{Absorption spectrum of three different initial nuclear states in the molecular system vs different parameters.} The top diagrams depict the corresponding electronic transitions: \textbf{(a)} Fock state $\ket{1}$; \textbf{(b)} thermal state; \textbf{(c)} damped vacuum state. The middle and bottom rows show the corresponding experimental results and theoretical expectations, respectively. For clarity, here we only show the typical spectrums with the corresponding parameter next to each plot. For both thermal state and damped vacuum state, $D=1$. Except for a reduction of experimental peak values, the experimental results show good agreement with theoretical expectation.}
\end{figure*}

%\begin{figure}
%   \includegraphics[scale=0.7]{other_cases.pdf}
%    \caption{\label{fig:other_cases}\textbf{Experimental and theoretical absorption spectrum of three different initial nuclear states in the molecular system.} The top diagrams depict the corresponding electronic transitions: \textbf{(a)} Fock state $\ket{1}$; \textbf{(b)} thermal state; \textbf{(c)} damped vacuum state. For clarity, here we only show the typical spectrums. Huang-Rhys parameter $D=1$ is fixed in \textbf{(b,e,h)} and \textbf{(c,f,i)}. \textbf{(a,d,g)} Fock state $\ket{1}$ with different Huang-Rhys parameter $D(=0.2,1.4,2.6,3.8)$. \textbf{(b,e,h)} Thermal equilibrium states with different temperatures described by thermally averaged occupation number of the harmonic vibrational mode $\bar{n} =1/(e^{\hbar\omega_0/kT}-1)$ where $k$ is Boltzmann constant$(\bar{n} =0,1.2,2.4,3.6)$.  \textbf{(c,f,i)} Damped vacuum states with different decay factors described by the characteristic time $\tau$($1/\tau\omega_0=0,0.2,0.4,0.6$). Excluding the reduction of experimental peak values which are dominantly caused by the qubit decoherence, the experimental results show good agreement with the theoretical predictions.}
%\end{figure}

\begin{figure*}
    \includegraphics[scale=1]{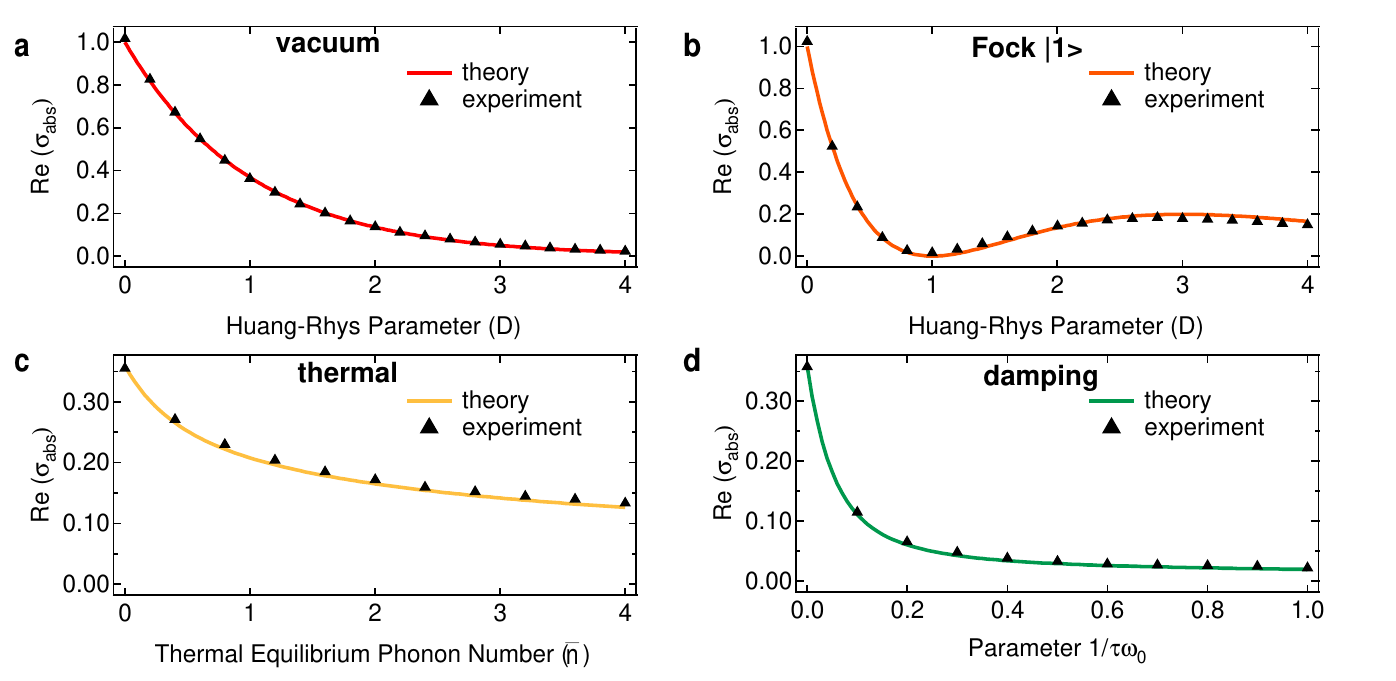}
    \caption{\label{fig:dec_0_point} \textbf{Progression of spectral peak at $\omega = \omega_{eg}$.} Peak values at $\omega = \omega_{eg}$ in Fig.~\ref{fig:zero_temperature} and Fig.~\ref{fig:other_cases} as a function of \textbf{(a)} $D$ for an initial vacuum state; \textbf{(b)} $D$ for a Fock state $\ket{1}$; \textbf{(c)} $\bar{n}$ for a thermal equilibrium state; \textbf{(d)} $1/\tau\omega_0$ for a damped vacuum state. Dots are experimental data after dividing a constant reduction factor $f$, in good agreement with theoretical expectation depicted in solid curves. $f=0.83$ for \textbf{a}, \textbf{b}, and \textbf{d} while $f=0.75$ for \textbf{c}. The smaller $f$ for the case of Fock state $\ket{1}$ is mainly due to the finite Fock state preparation fidelity $F=0.94$ while all other three cases start from a nearly perfect vacuum state. The standard deviation for each measured value is much less than 0.01 and not shown in the figure.}
\end{figure*}

In this work, we develop and demonstrate a quantum simulation approach for studying  molecular dynamics and absorption spectroscopy using a superconducting simulator. Besides simulating molecules in equilibrium, this approach of quantum simulation also allows us to obtain non-equilibrium molecular spectra that are not directly accessible under normal laboratory conditions. In addition, the problem of sampling the absorption spectra of molecules~\cite{Huh2014a} has been found to be related to the problem of Boson Sampling~\cite{Aaronson2013a}, which represents a promising approach to justify that quantum simulators cannot be simulated efficiently with any classical means. Our approach is complimentary with the existing approach~\cite{Huh2014a}, where the absorption spectra are obtained by sampling the transition probabilities for each pair of input-output Fock states. The key difference is that we focus on the dynamics of the phonons, instead of the structural shift due to the Duschinsky transformation~\cite{Peng2007}.

More specifically, our approach can be applied to obtaining the temporal correlation function of the electronic transition dipole~\cite{mukamel1999principles}, which yields the information about the absorption spectrum of the molecule, after applying the Fourier transformation. In our superconducting simulator, there are many adjustable control knobs for simulating the spectra for a variety of scenarios. In particular, we are able to simulate molecules in a wide range of values of the Huang-Rhys parameter $D$, which characterizes the electron-phonon coupling strength. 

In this work, we focus on the model approximating the electronic degrees of freedom by a two-level system (Fig.~1a). This model has been applied to study vibronic wavepacket dynamics, chemical reaction rate, Marcus theory for non-adiabatic electron transfer, etc. For molecular spectroscopy, the absorption spectra strongly depend on the initial state of the phonon degree of freedom in the manifold of the electronic ground state. In our experimental demonstration, we have performed simulations by preparing the phonon mode in pure Fock states, as well as simulations for a thermal state and a vacuum state with damping. In all cases, we are able to experimentally observe the progression of absorption peaks separated by the vibronic frequency, which is a characteristic feature of molecular spectrum due to vibronic transitions. This flexibility of our superconducting simulator makes it a useful tool for validating theoretical prediction when scaled up ({see Methods section}).

%\ls{The protocol of our simulator is shown in Fig.~\ref{fig:double_level}b. } 

The architecture of the superconducting simulator is constructed through a three-dimensional (3D) circuit quantum electrodynamics (QED) system~\cite{Wallraff,Paik}, where a ``vertical" transmon qubit is dispersively coupled to two 3D aluminum cavities for storage and readout, as shown in Fig.~1c. The qubit with a transition frequency $\omega_{eg}/2\pi=5.345$~GHz is used to model the electronic state $\{ \ket{g},\ket{e} \}$ of the molecule. The storage cavity (hereafter referred as the ``cavity" for simplicity) is used to model the quantization of the nuclear vibrational motion, i.e., phonons $\{ \ket{0}, \ket{1}, \ket{2}, ...\}$, with a frequency $\omega_0/2\pi=8.230$~GHz. Note that the energy gap of the qubit is comparable with that of the cavity frequency, i.e., $\omega_{eg} \sim \omega_0$. However, for a typical molecule, the phonon frequency is much smaller than that of the electronic excitation gap. Therefore, a direct analog molecular simulation with superconducting qubits is not feasible; such a challenge can be overcome by a digital approach of quantum simulation covered in this work. 

The working mechanism of our superconducting simulator is summarized as follows (see {Figs.~1b and 1c}). First, the qubit is initialized to the ground state~$\ket{g}$ while the phonons (cavity) are prepared in certain given state $\left| \psi  \right\rangle $ for the purpose of simulating the molecular system initially at different nuclear states. In our experiment, we have prepared different phonon states: (i) a vacuum state at zero temperature, (ii) a Fock state $\ket{1}$ at zero temperature, (iii) a thermal equilibrium state, and (iv) a vacuum state with damping. As an example, the pulse sequence for the case of a Fock state $\ket{1}$ is presented in the Supplementary Materials. The qubit is then through a classical microwave pulse turned into a superposition state $(\ket{g}+\ket{e})/\sqrt{2}$, after applying a $\pi/2$ rotation (a Hadamard transformation). 

{Next, a controlled-operation ${U_{\text{ctrl}}} $ is applied to the qubit-phonon system, which drives the evolution of the phonons only if the qubit is in $\ket{g}$, i.e., ${U_{\text{ctrl}}} = \left| g \right\rangle \left\langle g \right| \otimes U + \left| e \right\rangle \left\langle e \right| \otimes I$, where the unitary operator, $U (t) \equiv {e^{i{H_g}t}}{e^{ - i{H_e}t}}$, first evolves the phonons for a time interval $t$ with Hamiltonian $H_e$, followed by an inverse time evolution with $H_g$ for the same time interval.} {The operation $U$ can be simplified as follows: in the second quantized form, we have the Hamiltonian, ${H_e} = \omega_0 {b^\dag } b + \omega_{eg}$, describing a harmonic oscillator with an equilibrium position shifted by $d$ relative to ${H_g} = \omega_0 {a^\dag }a$, where $b = {\cal D} ( { - \tilde d} ) a {\cal D} ( {\tilde d} ) = a + {\tilde d}$ with $\tilde d = d\sqrt {m{\omega _0}/2\hbar}$, and ${\cal D}$ a displacement operator. Consequently, the operator $U$ can be implemented as a displacement operator, $U = {e^{-i \phi \left( t \right)}}{\cal D}(\tilde d({e^{i{\omega _0}t}} - 1))$, apart from a phase factor ${e^{-i\phi \left( t \right)}}$, where $\phi \left( t \right) \equiv {\omega _{eg}}t + {{\tilde d}^2}\sin {\omega _0}t$ (see Supplementary Materials for derivation details). 

Note that this phase factor cannot be ignored, as it yields a relative phase instead of global phase with $U_{\text{ctrl}}$. Experimentally, the phase $\phi$ is realized in the previous $\pi/2$ rotation as an azimuth angle in the $X$-$Y$ plane on the Bloch sphere (Fig.~1c). The controlled displacement operation ${\cal D}^{g}(\boldsymbol{\alpha})$, effective only when the qubit is at $\ket{g}$ state as indicated by an extra superscript $g$, is implemented by a broad selective pulse with a Gaussian envelope truncated to $4\sigma=1.34~\mu$s (Fig.~1c). Here the displacement vector ${\boldsymbol{\alpha}}=\tilde d({e^{i{\omega _0}t}} - 1)$. It is worth noting that the decoherence of the system during this long selective pulse lowers the subsequent qubit measurement contrast by a factor of about 0.83 compared to the ideal case.} 

Finally, as a result the dipole correlation function defined as ${C_{\mu \mu }}(t) = {\left\langle \psi  \right|U(t)\left| \psi  \right\rangle} $ is encoded in the off-diagonal elements of the reduced density matrix of the qubit, i.e., ${C_{\mu \mu }}\left( t \right) = \left\langle {{\sigma _x}} \right\rangle  + i\left\langle {{\sigma _y}} \right\rangle$. $\left\langle {{\sigma_y}} \right\rangle$ and $\left\langle {{\sigma_x}} \right\rangle$ of the qubit can be measured by applying an extra $\pi/2$ rotations along $X$ and $Y$ axis ($R_{X or Y}$) respectively followed by a $Z$-basis measurement. This general procedure is applicable for any initial state of the phonon, pure or mixed. The absorption spectrum ${\sigma _{{\text{abs}}}}$ is finally obtained by a Fourier transform of ${C_{\mu \mu }}(t)$.

We follow the above procedure to simulate the molecular system initially at a vacuum state and a Fock state $\ket{1}$ at zero temperature. However, in order to simulate molecular spectra with the phonon mode initialized in a thermal state, $\rho \equiv {e^{ - {\hbar\omega _0}{a^\dag }a/kT}}/{\text{Tr}}({e^{ - {\hbar\omega _0}{a^\dag }a/kT}})$, it is not practical to increase the physical temperature, as the performance of the experimental system would decrease significantly. To overcome this challenge, we can modify the above procedure at {Step 1}: instead of an equal superposition {(after a Hadamard gate)}, the qubit is initialized { to ${e^{ - i\phi (t) }}\sin{\frac{\gamma(t)}{2}}\ket{g}+ \cos{\frac{\gamma(t)}{2}}\ket{e}$, where the angle $\gamma(t)$ is chosen such that $\sin\gamma=e^{2 \tilde{d}^2\bar{n} (\cos \omega_0 t - 1)}$ and $\bar{n}=(e^{\hbar\omega_0 /kT}-1)^{-1}$ (see Supplementary Materials).} Similarly, for the case {of a vacuum state with damping}, we choose $\sin\gamma(t)={e^{ - t/\tau }}$, where $\tau$ is the characteristic time (also see Supplementary Materials). In both cases, following the same remaining procedure as described above, {one can obtain the correlation function ${C_{\mu \mu }^{thm}}\left( t \right)$ for an initial thermal state and the damped correlation function $C_{\mu \mu }^{{\text{damp}}}\left( t \right) = {e^{ - t/\tau }}{C_{\mu \mu }}\left( t \right)$, respectively.}

%We follow a similar procedure to simulate the molecular system initially at the vacuum state and a Fock state $\ket{1}$ at zero temperature. However, in order to simulate molecular spectra with line broadening, the above procedure has to be modified at {Step 1}: instead of an equal superposition {(after a Hadamard gate)}, the qubit is initialized {by $R_{\theta}(\gamma)$ to ${e^{ - i\theta (t) }}\sin{\frac{\gamma(t)}{2}}\ket{g}+ \cos{\frac{\gamma(t)}{2}}\ket{e}$, where the angle $\gamma(t)$ is chosen such that $\sin\gamma(t)={e^{ - t/\tau }}$ (see Supplementary Materials).} Following the same remaining procedure above, one can obtain the damped correlation function $C_{\mu \mu }^{{\text{damp}}}\left( t \right) = {e^{ - t/\tau }}{C_{\mu \mu }}\left( t \right)$.

%Our superconducting simulator can also be employed to simulate molecular spectra with the phonon mode initialized in a thermal state, ${\rho _\beta } \equiv {e^{ - \beta {\omega _0}{a^\dag }a}}/{\text{tr}}({e^{ - \beta {\omega _0}{a^\dag }a}})$. However, it is not practical to increase the physical temperature, as the performance of the experimental setup would decrease significantly. To overcome this challenge, {we use similar technique as for the case with line broadening, we initialize the qubit with $R_{\theta}(\gamma)$, where $\sin\gamma=e^{2 \tilde{d}^2\bar{n} (\cos (\omega_0 t) - 1)}$ and $\bar{n}=(e^{\hbar\omega_0 /kT}-1)^{-1}$ is the thermal average occupation number of the vibrational mode.}

 Our experimental results are as follows. {In our digital simulation, we have set $\Delta t=1$, $t_{max}=900$, $\omega_{eg}=\pi/5$, $\omega_0=\pi/90$. The spectrum lineshape of molecule illustrates the relative probability of electronic transition between different vibrational states in nuclear space.} In Fig.~\ref{fig:zero_temperature}, we present the progression of absorption peaks for the case where the phonon state is initialized at vacuum and at zero temperature, {$\left| \psi \right\rangle  = \left| 0 \right\rangle$}, for various Huang-Rhys parameter $D={\tilde d}^2$. When $D=0$, there is only a sharp peak located at the frequency $\omega = \omega_{eg}$. This case represents the limit where the electronic transition and the nuclear motion are decoupled. In other words, the molecule is essentially the same as a two-level atom, as far as the spectrum is concerned. When $D$ is increased from zero to, e.g., $D=1$, several peaks emerge, and these peaks are equally spaced by the phonon frequency $\omega_0$. When $D$ is increased further to $D=4$, we can observe more equally-spaced peaks. However, the amplitude of the direct transition at $\omega = \omega_{eg}$ is no longer the largest. In all experimental trials, except for a reduction factor $f=0.83$ mainly due to the qubit decoherence, the spectral peaks are in good agreement with the expected Poisson distribution (see Methods).

%\ls{Furthermore, all the spectrum peaks are consistent with the expected Poisson distribution~\cite{mukamel1999principles}.}

%These results are consistent with the approximation of the spectrum envelope based on a Gaussian profile, i.e., ${\sigma _{{\text{abs}}}}\left( \omega  \right) \propto {e^{ - {{(\omega  - {\omega _{eg}} - D{\omega _0})}^2}/2D\omega _0^2}}$, which has a maximum at $\omega  - {\omega _{eg}} = D{\omega _0}$.

The absorption spectrum of the other three different initial nuclear states in the molecular system are shown in {Fig.~\ref{fig:other_cases}: (i) Fock state $\ket{1}$ with different $D$; (ii) thermal equilibrium state at different temperatures characterized by the occupation number $\bar{n}$; and (iii) damped vacuum state with different dissipation rates described by the characteristic time $\tau$. The corresponding electronic transitions for each case have been depicted in the top diagrams of Fig.~\ref{fig:other_cases}. For clarity, we only show the typical spectra. Except for a reduction of experimental peak values, the experimental results show good agreement with  theoretical expectation.}

%The spectrum lineshape of molecule illustrates the relative probability of electronic transition between different vibrational states in nuclear space. According to Franck$-$Condon principle, one of the reasons that lead to the modification of vibrational states between two electronic surfaces is the coupling between electronic and nuclear system which depicted by Huang-Rhys parameter $D$. Thus an interesting problem is, what is the relationship between vibrational invariance[term named by me, is there other better name??] and $D$(or other variables, such as temperature or damping factor). Here vibrational invariance that depicted by the peak of $\omega = \omega_{eg}$ in the spectrum illustrates the probability that nuclear vibrational states do not modify after electronic transition. To demonstrate the advantage of our quantum simulation method, we report the modification of vibrational invariance probability with different parameters in Fig ~\ref{fig:dec_0_point}. The result illustrates that, in our molecular model of two-level electronic harmonic oscillators potentials, the invariance probability of nuclear vibrational states after electronic transition decreases with environment becoming hotter(Fig ~\ref{fig:dec_0_point} \textbf{b}) or more dissipative(Fig ~\ref{fig:dec_0_point} \textbf{d}). In the low temperature(i.e $\ket{0}$) without dissipation, the probability decreases as $D$ grows(Fig ~\ref{fig:dec_0_point} \textbf{a}). However, the theorem fails if the nuclear system are not initially in thermal equilibrium(Fig ~\ref{fig:dec_0_point} \textbf{c}).

{One of the key features of our quantum simulator is that the parameters, such as Huang-Rhys parameter $D$, can be varied continuously. To better illustrate the progression of the spectrum in Fig.~\ref{fig:zero_temperature} and Fig.~\ref{fig:other_cases} as a function of various parameters, we present the peak values at $\omega = \omega_{eg}$ as an example in Fig.~\ref{fig:dec_0_point}. Dots are experimental data by our quantum simulator while the solid curves represent theoretical expectation. After taking into account the reduction of experimental peak values, again mainly due to the system decoherence, by dividing a constant reduction factor $f$ ($f=0.83$ for Figs.~\ref{fig:dec_0_point}a, \ref{fig:dec_0_point}c, and \ref{fig:dec_0_point}d; $f=0.75$ for Fig.~\ref{fig:dec_0_point}b), the experimental results are in good agreement with theoretical expectations. The smaller $f$ for the case of Fock state $\ket{1}$ is mainly due to the finite Fock state preparation fidelity $F=0.94$ (measured Wigner function shown in Supplementary Materials) while all other three cases start from a nearly perfect vacuum state.}

%\section{Conclusion}
To conclude, we demonstrated experimentally a new method to simulate electronic absorption spectra of a molecule, where the nuclear vibrational states may or may not be in thermal equilibrium. Our quantum simulator is based on a superconducting circuit QED architecture with flexible parameter tunability. The simulation results indicate that the resulting molecular spectra are in good agreement with theoretical expectation. Finally, we note that this method can be readily extended to other quantum simulation platform, including photonic~\cite{aspuru2012photonic} or trapped-ion~\cite{kim2010quantum} systems. Therefore, our experiment represents the beginning of a new approach of predicting molecular spectroscopy using quantum  simulators.

%Furthermore, the method and experimental setup are both scalable for a further development.   

%\begin{methods}

\subsection{Device parameters.}
The transmon qubit has an energy-relaxation time $T_1=13~\mu$s and a pure dephasing time $T_{\phi}=16~\mu$s. The storage cavity has a lifetime $\tau_0=80~\mu$s. The readout cavity has a transition frequency $\omega_m/2\pi=7.291$~GHz and a lifetime $\tau_r=42$~ns. Together with a Josephson parametric amplifier~\cite{Hatridge2011,Roy2015} operating in a double-pumped mode~\cite{Kamal2009,Murch2013}, the fast readout cavity is used for a high fidelity and quantum non-demolition detection of the qubit state  (see Supplementary Materials for details). Experimental setup details can also be found in Ref.~\cite{Liu2016}. The qubit-state-dependent frequency shift of the storage cavity is $\chi_{qs}/2\pi=-1.44$~MHz, allowing for the qubit-controlled operation on the cavity state as used in our experiment.

\subsection{Molecular Hamiltonian.}
%We first summarize the molecular physics involved in our work, before presenting the experimental methods and results. 
Under the standard Born-Oppenheimer framework, the Hamiltonian $H_{\rm mol}$ of a molecule depends on the {nuclear configuration (i.e., position coordinates) $\bm q$ as parameters}, ${{H_{\text{mol}}(\bm r, \bm q)}} = {K_{\text{e}}}  + {U_{{\text{ee}}}}\left( {\bm r} \right) + {U_{\text{eN}}}\left( {{{\bm r},{\bm q}}} \right)$, where ${K_{\text{e}}}$ is the kinetic-energy term for the electrons, ${U_{{\text{ee}}}}\left( {\bm r} \right)$ and ${U_{\text{eN}}}\left( {{{\bm r},{\bm q}}} \right)$ are the electron-electron interaction term and electron-nuclei interaction term respectively. In the low-energy sector, the molecule typically contains an electronic ground state $\ket{g}$ and an excited state $\ket{e}$, where the molecular Hamiltonian becomes: 
\begin{equation}
H_{\rm mol} ({\bm q}) = H_g(\bm{q}) \ket{g}  \bra{g}  + H_e(\bm{q}) \ket{e} \bra{e} \ ,
\end{equation}
with $H_g = K_{\rm N} +V_g(\bm{q})$ and $H_e = K_{\rm N} +V_e(\bm{q})$. Here $K_{\rm N}$ is the nuclear kinetic energy, $V_g(\bm{q})$ and $V_e(\bm{q})$ are the potential energies, which are typically approximated as harmonic functions (Fig.~\ref{fig:double_level}a), i.e., ${H_g} = \tfrac{1}{{2m}}{p^2} + \tfrac{{m{\omega_0 ^2}}}{2}{q^2}$ and ${H_e} = \tfrac{1}{{2m}}{p^2} + \tfrac{{m{\omega_0 ^2}}}{2}{(q - d)^2} + {\hbar\omega _{eg}}$. Here $\omega_{eg}$ is the electronic gap between the minima of both potentials (i.e., 0-0 energy splitting).   

\subsection{Franck-Condon approximation.}
The coupling strength between the electronic transition and the nuclear motion is characterized by the Huang-Rhys parameter, $D=\tilde d^2$, where $\tilde d = d\sqrt {m{\omega _0}/2\hbar}$. Similarly, the electronic transition dipole operator is given by $\mu(\bm{q})=\mu_{eg}(\bm{q})\ket{e}\bra{g} + \mu_{ge}(\bm{q})\ket{g}\bra{e}$. However, the dependence of electronic transition moment on nuclear is usually insensitive to the nuclear motion; one can therefore approximate (known as Condon approximation) it with a constant, i.e., $\mu_{eg}(\bm{q}) = \mu_{ge}(\bm{q}) = 1$ for simplicity. 

\subsection{Absorption lineshape.}
The absorption line shape, $\sigma _{{\text{abs}}}\left( \omega  \right) = \int_{ -\infty }^\infty  {dt} \ {e^{i\omega t}} \ {C_{\mu \mu }}\left( t \right)$, can be obtained by the Fourier transform of the dipole correlation function ${C_{\mu \mu }}\left( t \right) \equiv \left\langle {\mu \left( t \right)\mu \left( 0 \right)} \right\rangle$, where $\mu \left( t \right) = {e^{i{H_{{\text{mol}}}}t/\hbar}}{\mu(0)} \ {e^{ - i{H_{{\text{mol}}}}t/\hbar}}$.  In order to mimic the effects on the molecular spectra due to the influence of the environment~\cite{kroto1992molecular}, one can append a damping factor ${e^{ - t/\tau }}$ to the above correlation function, i.e., $C_{\mu \mu }^{{\text{damp}}}\left( t \right) = {e^{ - t/\tau }}{C_{\mu \mu }}\left( t \right)$, which yields a spectrum with line broadening  $\sigma _{{\text{abs}}}^{{\text{damp}}}\left( \omega  \right)$. Our main task is to apply our superconducting simulator to obtain the correlation functions for the molecules to be simulated. For example, if the initial state is a vacuum state, the correlation function $C_{\mu\mu}\left( t \right) = e^{-i\omega_{eg}t}e^{D(e^{-i\omega_0 t}-1)}$. The absorption lineshape is $\sigma _{{\text{abs}}}\left( \omega  \right) = e^{-D}\int_{ -\infty }^\infty  {dt} \ e^{i\omega t} \ e^{-i\omega_{eg}t} e^{De^{-i \omega_0 t}}$. By expanding $e^{De^{-i \omega_0 t}} = \sum_{j = 0}^{ \infty }\frac{1}{ j! } \left( D e^{-i \omega_0 t} \right)^j$, the lineshape becomes $\sigma _{{\text{abs}}}\left( \omega  \right) = e^{ -D } \sum_{j = 0}^{ \infty }\frac{D^j}{ j! } \delta\left( \omega - \omega_{eg} - j\omega_0 \right)$. Thus the spectral peaks are separated by $\omega_0$ with a Poisson distribution of intensities.

%Thus the amplitude of absorption peaks are given by $\left| \left\langle 0|j \right\rangle \right|^2 = e^{-D}\frac{1}{j!}D^j$.}

\subsection{Scalability.}
Our approach can be scaled up for molecules with multiple vibronic modes. In this case, the dipole correlation function comes from the contributions of the individual modes, i.e., for $n$ modes, ${C_{\mu \mu }}\left( t \right) = {\left| {{\mu _{eg}}} \right|^2}{e^{ - i\left( {{E_e} - {E_g}} \right)t/\hbar }}{F_n}\left( t \right)$, where ${F_n}\left( t \right) = {\rm Tr}\left( {{e^{i{H_g}t/\hbar }}{e^{ - i{H_e}t/\hbar }}{\rho _1} \otimes {\rho _2} \cdots  \otimes {\rho _n}} \right)$. In other words, the superconducting qubit needs to be coupled with multiple cavity modes. This direction has been realized experimentally~\cite{Wang2016b}. There, a superconducting qubit is coupled to two cavity modes to realize an entangled pair of single-cavity cat states. With a similar geometry, the superconducting qubit can easily be extended to couple to more cavity modes.

%\end{methods}

\begin{acknowledgments}
We thank R.~Vijay and his group for the help with the parametric amplifier. This work was supported by the National Natural Science Foundation of China under Grant No. 11474177, the Ministry of Science and the Ministry of Education of China through its grant to Tsinghua University, the Major State Basic Research Development Program of China under Grant No.2012CB921601, and the 1000 Youth Fellowship program in China.

\end{acknowledgments} 

%

%% Here is the endmatter stuff: Supplementary Info, etc.
%% Use \item's to separate, default label is "Acknowledgements"

\end{document}

\title{Supplementary Information``Simulating Molecular Spectroscopy with Circuit Quantum Electrodynamics"}
\author{L. H$^*$}
\author{Y. C. Ma$^*$}
\author{Y. Xu}
\author{W. Wang}
\author{Y. Ma}
\author{K. Liu}
\affiliation{Center for Quantum Information, Institute for Interdisciplinary Information Sciences, Tsinghua University, Beijing 100084, China}
\author{M.-H. Yung$^\dagger$}
\email{yung@sustc.edu.cn}
\affiliation{Institute for Quantum Science and Engineering and Department of Physics, South University of Science and Technology of China, Shenzhen 518055, China}
\author{L.~Sun$^\dagger$}
\email{luyansun@tsinghua.edu.cn}
\affiliation{Center for Quantum Information, Institute for Interdisciplinary Information Sciences, Tsinghua University, Beijing 100084, China}
% It is always \today, today,
             %  but any date may be explicitly specified
\maketitle

\tableofcontents

\newpage

{\section{Key ideas of the molecular spectroscopy simulation}}
The main idea of this work is related to the problem of simulating the absorption spectrum of molecules associated with the Huang-Rhys factor $D$. The molecular spectrum can be obtained by applying the Fourier transformation on the temporal correlation function of an electronic transition dipole operator. Therefore the key part of the experiment involves a simulation of the correlation function through a quantum circuit.

As shown in Fig.~1b in the main text, the core of the quantum circuit consists of three components:

\begin{enumerate}
\item Hadamard gate,
\begin{equation}
    H \equiv \frac{1}{\sqrt{2}}(\ket{g}+\ket{e})\bra{g}+\frac{1}{\sqrt{2}}(\ket{g}-\ket{e})\bra{e},
\end{equation}
applied to the ancilla qubit at the beginning of our quantum circuit. Here $\ket{g}$ represents the ground state and $\ket{e}$ the excited state of the qubit. 

\item A controlled unitary operation
\begin{equation}
U_{\text{ctrl}}=\ket{g}\bra{g}\otimes U+\ket{e}\bra{e}\otimes I,
\end{equation}
where $U\equiv e^{iH_gt/\hbar}e^{-iH_et/\hbar}$ is the $``$Forward and Reversal" time-evolution operator and  $I$ is the identity operator.

\item $\sigma_y$ and $\sigma_x$ measurements through $\pi/2$ rotations along $X$ and $Y$ axis:
\begin{eqnarray}\label{eq:xy_rot}
     R_X &\equiv& \frac{1}{\sqrt{2}}(\ket{g}-i\ket{e})\bra{g}+\frac{1}{\sqrt{2}}(\ket{e}-i\ket{g})\bra{e},\\
     R_Y &\equiv& \frac{1}{\sqrt{2}}(\ket{g}+\ket{e})\bra{g}+\frac{1}{\sqrt{2}}(\ket{e}-\ket{g})\bra{e}.
\end{eqnarray}
%Measure $\sigma_z$ following by $x$ and $y$ Rotation gate.
Similar to DQC1, We get the real and imaginary part of the correlation function ${C_{\mu \mu }}(t) = {\left\langle \psi  \right|U(t)\left| \psi  \right\rangle} = {\left\langle \sigma_x\right\rangle} + i{\left\langle \sigma_y\right\rangle}$, where $\ket{\psi}$ is the phonon initial state. Then the absorption spectrum of the molecule can be obtained by a Fourier transformation of $C_{\mu\mu}(t)$.
\end{enumerate}

\section{Temporal Correlation function}
Under the Condon approximation, the dipole operator is given by, 
\begin{equation}
\mu = \ket{e}\bra{g} + \ket{g}\bra{e} = \sigma_x \ ,
\end{equation}
where $\ket{g}$ and $\ket{e}$ represent the ground and excited electronic wave function. Given a molecular Hamiltonian,
\begin{equation}
H = H_g \ket{g}  \bra{g}  + H_e \ket{e} \bra{e} \ ,
\end{equation}
the time-correlation function $C_{\mu\mu}(t)$ for the dipole operator is given by 
\begin{equation}
C_{\mu\mu}(t) \equiv {\rm Tr}(\bra{g}\sigma_x(t)\sigma_x(t=0)\ket{g}) \ ,
\end{equation}
where ${\sigma _x}\left( t \right) = {e^{iHt/\hbar }}{\sigma _x}{e^{ - iHt/\hbar }}$, and ${\sigma _x}\left( {t = 0} \right) = {\sigma _x}$. Since ${\sigma _x}\left| g \right\rangle  = \left| e \right\rangle $, we can write 
\begin{eqnarray}
C_{\mu\mu}(t) &=& {\rm Tr}(\bra{g}e^{iH_g/\hbar\ket{g}\bra{g}t}\sigma_xe^{-iH_e/\hbar\ket{e}\bra{e}t}\ket{e}) \nonumber \\
&=& {\rm Tr}(e^{iH_gt/\hbar}e^{-iH_et/\hbar}).
\end{eqnarray}
Here ${{\rm Tr}}(\cdot)$ represents the expectation of the operator in the nuclear space. For simplicity, ${\rm Tr}(\cdot)$ is replaced by $\braket{\cdot}$ hereafter. 

\section{Harmonic approximation}
In terms of the creation and annihilation operators, 
\begin{equation}
p = i\sqrt{m\hbar\omega_0/2}(a^{\dagger}-a) \quad {\rm and} \quad q = \sqrt{\hbar/2m\omega_0}(a+a^{\dagger}) \ .
\end{equation}
The Hamiltonians of the ground and excited electronic states are given by
\begin{eqnarray}
H_g &=& \hbar\omega_0 a^{\dagger}a \ , \\
H_e &=& \hbar\omega_{eg} + \hbar\omega_0 a^{\dagger}a + \tilde{d}(a+a^{\dagger}) + \hbar\omega_0\tilde{d}^2 \nonumber \\
&=& \hbar\omega_{eg} + \hbar\omega_0 {\cal D}(-\tilde{d})a^{\dagger}a{\cal D}(\tilde{d}) \ .
\end{eqnarray}
Here ${\cal D}(\tilde{d}) = e^{\tilde{d}a^{\dagger}-\tilde{d}^{\star}a}$ is the displacement operator for a single mode. The Huang-Rhys parameter $D$ is defined as $D=\tilde d^2$. 

With ${\cal D}(-\tilde{d})a{\cal D}(\tilde{d}) = a + \tilde{d}$, we have
\begin{eqnarray}
C_{\mu\mu}(t) &=& e^{-i\omega_{eg}t}\braket{e^{i\omega_0 a^{\dagger}at}{\cal D}(-\tilde{d})e^{-i\omega_0 a^{\dagger}at}{\cal D}(\tilde{d})}\nonumber\\
&=& e^{-i\omega_{eg}t}\braket{e^{\tilde{d}(a^{\dagger}e^{i\omega_0 t}-ae^{-i\omega_0 t})}{\cal D}(\tilde{d})} \nonumber \\
&=& e^{-i\omega_{eg}t}\braket{{\cal D}(\tilde{d}e^{i\omega_0 t}){\cal D}(-\tilde{d})} \nonumber \\
&=& e^{-i\omega_{eg}t-i\tilde{d}^2\sin\omega_0 t}\braket{{\cal D}(\tilde{d}(e^{i\omega_0 t}-1))} \ .
\end{eqnarray}

\section{Correlation function of Fock states}
The states in the nuclear space can be written as linear superpositions of $\ket{n},n=0,1,2,\cdots$. Note that
\begin{eqnarray*}
e^{-\alpha^* a}\ket{n} =
\ket{n}+\sqrt{n}(-\alpha^*)\ket{n-1} +
\sqrt{n(n-1)}(\alpha^*)^2/2!\ket{n-2} + \cdots +  \sqrt{n!}(-\alpha^*)^n/(n!)^2\ket{0}.
\end{eqnarray*}
Therefore, for Fock states, if we define $\alpha=\tilde{d}(e^{i\omega_0 t}-1)$, we have
\begin{eqnarray}
C_{\mu\mu}(t) 
&=& e^{-i\omega_{eg}t-i\tilde{d}^2\sin\omega_0 t}\bra{n}{\cal D}(\alpha)\ket{n} \nonumber \\
&=& e^{-i\omega_{eg}t-i\tilde{d}^2\sin\omega_0 t}\bra{n}e^{-\frac{|\alpha|^2}{2}}e^{\alpha a^{\dagger}}e^{-\alpha^* a}\ket{n} \nonumber \\
&=&e^{-i\omega_{eg}t-i\tilde{d}^2\sin\omega_0 t} e^{-\frac{|\alpha|^2}{2}}\sum_{j=0}^{n}(-1)^j\frac{n(n-1)\cdots(n-j)}{(j!)^2}|\alpha|^{2j}
\end{eqnarray}
For $\ket{n} = \ket{1}$,
\begin{eqnarray}
C_{\mu\mu}(t) = e^{-i\omega_{eg}t}e^{\tilde{d}^2(e^{-i\omega_0 t}-1)}(1-4\tilde{d}^2\sin^2(\omega_0 t/2)).
\end{eqnarray}
For $\ket{n} = \ket{0}$ (the equilibrium state at zero temperature),
\begin{eqnarray}
C_{\mu\mu}(t) = e^{-i\omega_{eg}t}e^{\tilde{d}^2(e^{-i\omega_0 t}-1)}. 
\end{eqnarray}

%For $\ket{n} = \ket{2}$
%\begin{equation}
%C_{\mu\mu}(t) = e^{-i\omega_{eg}t}e^{\tilde{d}^2(e^{-i\omega_0 t}-1)}(1-8\tilde{d}^2\sin^2(\omega_0 t/2)+8\tilde{d}^4\sin^4(\omega_0 t/2))
%\end{equation}

%\section{Correlation function for coherent state}

%Next we calculate the correlation function of coherence state, which can be prepared directly by circuit QED.
%\begin{eqnarray*}
%C_{\mu\mu}(t) &=& e^{-i\omega_{eg}t-i\tilde{d}^2\sin\omega_0 t}\bra{\alpha}{\cal D}(\tilde{d}(e^{i\omega_0 t}-1))\ket{\alpha}\\
%&=& e^{-i\omega_{eg}t-i\tilde{d}^2\sin\omega_0 t}\bra{0}{\cal D}(-\alpha){\cal D}(\tilde{d}(e^{i\omega_0 t}-1)){\cal D}(\alpha)\ket{0}\\
%&=& e^{-i\omega_{eg}t+2i\tilde{d}\alpha\sin(\omega_0 t)+\tilde{d}^2(e^{-i\omega_0 t}-1)}
% \end{eqnarray*}

\section{Evolution Process in experiment}\label{sec:exp}
Our quantum simulator simulates the absorption spectra of a molecule in which electronic states (initialized at the ground state $\ket{g}$) coupled with a single nuclear mode (denoted by $\ket{\psi}$). In the controlled unitary gate $U_{\text{ctrl}} = \left| g \right\rangle \left\langle g \right| \otimes U + \left| e \right\rangle \left\langle e \right| \otimes I$, we have
\begin{eqnarray}
U\equiv e^{iH_gt/\hbar}e^{-iH_et/\hbar}=e^{-i\omega_{eg}t-i\tilde{d}^2\sin\omega_0 t}{\cal D}(\tilde{d}(e^{i\omega_0 t}-1)).
\end{eqnarray}
 The standard state evolution steps of our quantum simulator are 
\begin{eqnarray*}
\ket{g}\ket{\psi} &\xrightarrow{\text{Hadamard}}& \frac{1}{\sqrt{2}} (\ket{e}\ket{\psi}+ \ket{g}\ket{\psi})\\
&\xrightarrow{U_{\text{ctrl}} \text{ Gate}}& \frac{1}{\sqrt{2}}(\ket{e}\ket{\psi}+ e^{-i\omega_{eg}t-i\tilde{d}^2\sin \omega_0 t}\ket{g}{\cal D} (\tilde{d} (e^{i\omega_0 t} - 1))\ket{\psi}) \ .
\end{eqnarray*}
By measuring the expectation of Pauli operators $\sigma_x$ and $\sigma_y$, we get
\begin{eqnarray}
\braket{\sigma_x} &=& Re(e^{- i \omega_{eg} t - i \tilde{d}^2 \sin
	\omega_0 t} {\cal D} (\tilde{d} (e^{i \omega_0 t} - 1))) \ ,\\
\braket{\sigma_y} &=& Im(e^{- i \omega_{eg} t - i \tilde{d}^2 \sin
	\omega_0 t} {\cal D} (\tilde{d} (e^{i \omega_0 t} - 1))) \ .
\end{eqnarray}

In order to simulate molecular spectra with the phonon mode initialized in a thermal state, $\rho \equiv {e^{ - {\hbar\omega _0}{a^\dag }a/kT}}/{\text{Tr}}({e^{ - {\hbar\omega _0}{a^\dag }a/kT}})$, it is not practical to increase the physical temperature, as the performance of the experimental system would decrease significantly. To overcome this challenge, we can modify the above procedure at the first step: instead of an equal superposition {(after a Hadamard gate)}, the qubit is initialized to $e^{ - i\phi (t)}\sin{\frac{\gamma(t)}{2}}\ket{g}+ \cos{\frac{\gamma(t)}{2}}\ket{e}$, where $\phi \left( t \right) \equiv {\omega _{eg}}t + {{\tilde d}^2}\sin {\omega _0}t$, the angle $\gamma(t)$ is chosen such that $\sin\gamma=e^{2 \tilde{d}^2\bar{n} (\cos (\omega_0 t - 1)}$, and $\bar{n}=(e^{\hbar\omega_0 /kT}-1)^{-1}$.

%the first Hadamard gate is replaced by $R_{\theta} (\frac{\gamma}/2)$ to illustrate thermal equilibrium state, that is

%\begin{eqnarray*}
%\ket{g}\ket{0} &\xrightarrow& \frac{1}{\sqrt{2}} \left( \cos \alpha \ket{e}\ket{0} + e^{- i \omega_{eg} t - i \tilde{d}^2 \sin\theta}\sin\alpha \ket{g}\ket{0} \right)
%\end{eqnarray*}

Then we can re-calculate $\braket{\sigma_x}$ and $\braket{\sigma_y}$ to get $C_{\mu\mu}^{thm}(t)={\left\langle \sigma_x\right\rangle} + i{\left\langle \sigma_y\right\rangle}$ as follows.
\begin{eqnarray}
\braket{\sigma_x} &=& \sin \gamma Re(e^{- i \omega_{eg} t - i\tilde{d}^2 \sin \omega_0 t - | \tilde{d} (e^{i \omega_0 t} - 1) |^2 / 2}) \nonumber \\
%&=& \sin (2 \alpha) Re(e^{- i \omega_{eg} t - i \tilde{d}^2 \sin\theta - \tilde{d}^2 (1 - \cos \theta)})\\
&=& \sin \gamma Re(e^{- i \omega_{eg} t + \tilde{d}^2 (e^{-i \omega_0 t} - 1)}) \nonumber \\
&=& Re(e^{- i \omega_{eg} t + 2 \tilde{d}^2 \bar{n} (\cos \omega_0 t- 1) + \tilde{d}^2 (e^{- i \omega_0 t} - 1)}) \nonumber \\
&=& Re (e^{- i \omega_{eg} t + \tilde{d}^2 [(\bar{n} +1) (e^{- i \omega_0 t} - 1) + \bar{n} (e^{i \omega_0 t} - 1)]})\\
\braket{\sigma_y} &=& Im(e^{- i \omega_{eg} t + \tilde{d}^2 [(\bar{n} +1) (e^{- i \omega_0 t} - 1) + \bar{n} (e^{i \omega_0 t} - 1)]})
\end{eqnarray}

This result is consistent with the correlation function for a real thermal equilibrium state (also see \cite{mukamel1999principles})
\begin{eqnarray}
C_{\mu\mu}^{thm}(t) = e^{- i \omega_{eg} t + \tilde{d}^2 [(\bar{n} +1) (e^{- i \omega_0 t} - 1) + \bar{n} (e^{i \omega_0 t} - 1)]}.
\end{eqnarray}

\section{Simulating correlation function with damping}
The method to mimic the influence of the environment is similar to the case of a thermal equilibrium state except for $\sin\gamma(t) = e^{- t / \tau}$, where $\tau$ is the characteristic time which describes the decay time of the correlation function. This method works for any nuclear state $\ket{\psi}$
\begin{eqnarray}
\ket{g}\ket{\psi} &\Rightarrow& \cos \frac{\gamma(t)}{2} \ket{e}\ket{\psi} + e^{- i \omega_{eg} t - i \tilde{d}^2 \sin\omega_0 t}\sin\frac{\gamma(t)}{2} \ket{g}\ket{\psi} 
\end{eqnarray}

It is easy to verify that
\begin{equation}
C_{\mu\mu}^{\text{damp}}(t) = \braket{\sigma_x} + i \braket{\sigma_y} = e^{- t / \tau} C_{\mu \mu} (t)
\end{equation}
where $C_{\mu\mu}^{\text{damp}}(t)~(C_{\mu \mu} (t))$ is the correlation function with (without) damping.

{\section{Device and Readout properties}}
The transmon qubit in our experiment is fabricated using the standard Dolan technique~\cite{Dolan} with a double-angle evaporation of aluminum after a single electron-beam lithography step on a $c$-plane sapphire substrate. The experiment is performed in a cryogen-free dilution refrigerator with a base temperature of about 10~mK. A Josephson parametric amplifier (JPA)~\cite{Hatridge,Roy} is connected to the output of the readout cavity at the base temperature as the first stage of amplification before the high-electron-mobility-transistor amplifier at 4K. The JPA is operated in a pulsed double-pumped mode~\cite{Kamal,Murch} to minimize pump leakage to the readout cavity. The readout pulse is calibrated to contain only a few photons (see the calibration session below), enough for a high-fidelity single-shot readout of the qubit state. The schematic of the measurement setup can be found in Ref.~\onlinecite{Liu2016}. 

\begin{figure*}
\centering
\includegraphics{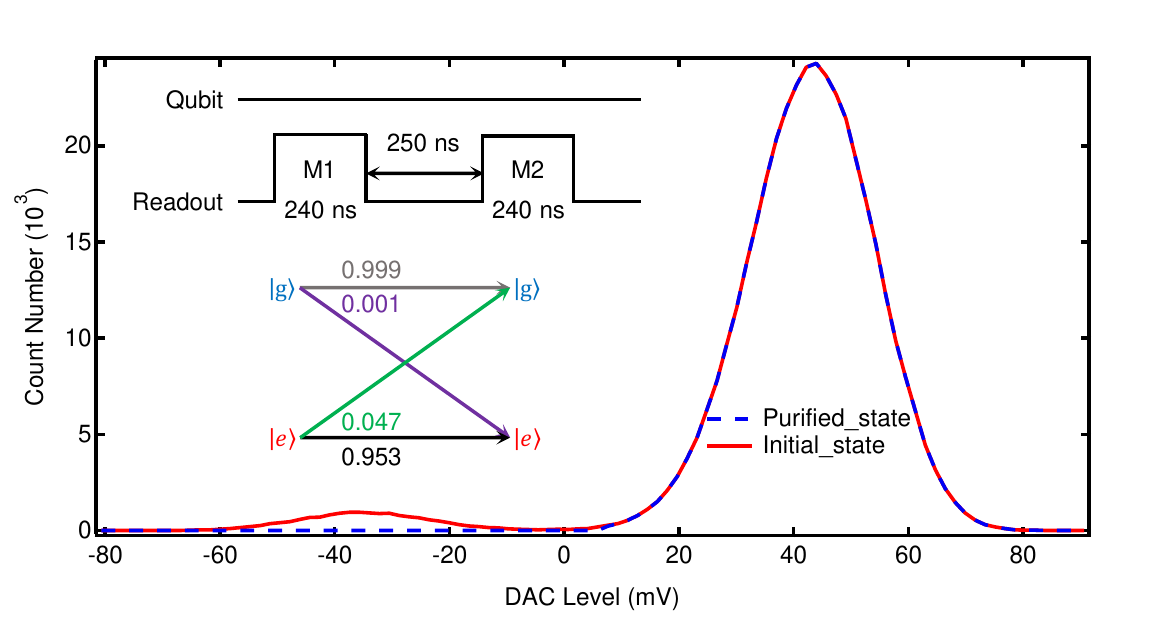}
\caption{Qubit readout property. Histogram of the readout with (the blue dashed line) and without (the red solid line) purification. The ground state $\ket{g}$ and the excited state $\ket{e}$ are well-separated and a threshold $V_{th}=0$ is chosen to digitize the readout result. Without purification, $\ket{e}$ takes up to $3.9\%$ occupation in the steady state, presumably due to stray infrared photons or other background noise leaking into the cavity. Inset is the basic qubit readout matrix and the corresponding experimental protocol with the storage cavity left in vacuum. After the purification to $\ket{g}$, the subsequent qubit measurement has 99.9\% probability of measuring $\ket{g}$ again, demonstrating the high QND nature of the qubit readout. If $\ket{e}$ state is post-selected, the subsequent qubit measurement has 95.3\% probability of measuring $\ket{e}$ again. The errors dominantly come from the $T_1$ process during the waiting time after the initialization measurement (250~ns) and during the readout time (240~ns).}
\label{fig:readoutproperty}
\end{figure*}

The readout property of the qubit is shown in Fig.~\ref{fig:readoutproperty}. The readout histogram is clearly bimodal and well separated. A threshold $V_{th}=0$ is chosen to digitize the readout signal to $+1$ and $-1$ for the ground state $\ket{g}$ and the excited state $\ket{e}$, respectively. The qubit has an excited state population of $3.9\%$ in the steady state, presumably due to stray infrared photons or other background noise leaking into the cavity, although the exact source remains unknown. We use measurement-based post-selection to purify the qubit to $\ket{g}$ state. This requires the measurement to be high quantum non-demolition (QND). Inset of Fig.~\ref{fig:readoutproperty} shows the qubit readout matrix with the cavity left in vacuum and the corresponding experiment sequence. After the first measurement and post-selection of $\ket{g}$, the subsequent qubit measurement shows that the probability of the qubit being populated in $\ket{g}$ is as high as 0.999 (dashed histogram), demonstrating the high QND property of the qubit readout. If we post-select the qubit at $\ket{e}$ state, the subsequent measurement finds that the qubit remains in $\ket{e}$ with a fidelity of 0.953. These errors dominantly come from the $T_1$ process during the waiting time after the initialization measurement (250~ns) and during the readout time (240~ns). 

\begin{figure*}
\centering
\includegraphics{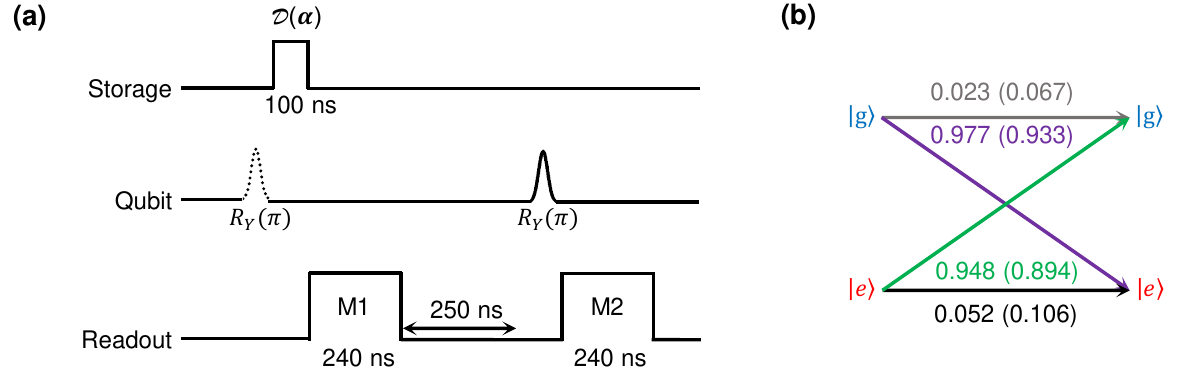}
\caption{Properties of non-selective qubit $\pi$ pulses with a coherent state $\ket{\alpha}$ in the cavity. \textbf{(a)} Calibration protocol. The absence or presence of a $\pi$ pulse at the beginning together with M1 is used to prepare the qubit's initial state $\ket{g}$ or $\ket{e}$ by a post-selection. \textbf{(b)} Readout matrix. The numbers outside the parenthesis correspond to a $\pi$ pulse with $\sigma=6$~ns for $\alpha=1$ ($\bar{n}=1$). The numbers inside the parenthesis correspond to a $\pi$ pulse with $\sigma=2$~ns for $\alpha=4$ ($\bar{n}=16$).}
\label{fig:pi_pulse_property}
\end{figure*}

\begin{figure*}
\centering
\includegraphics{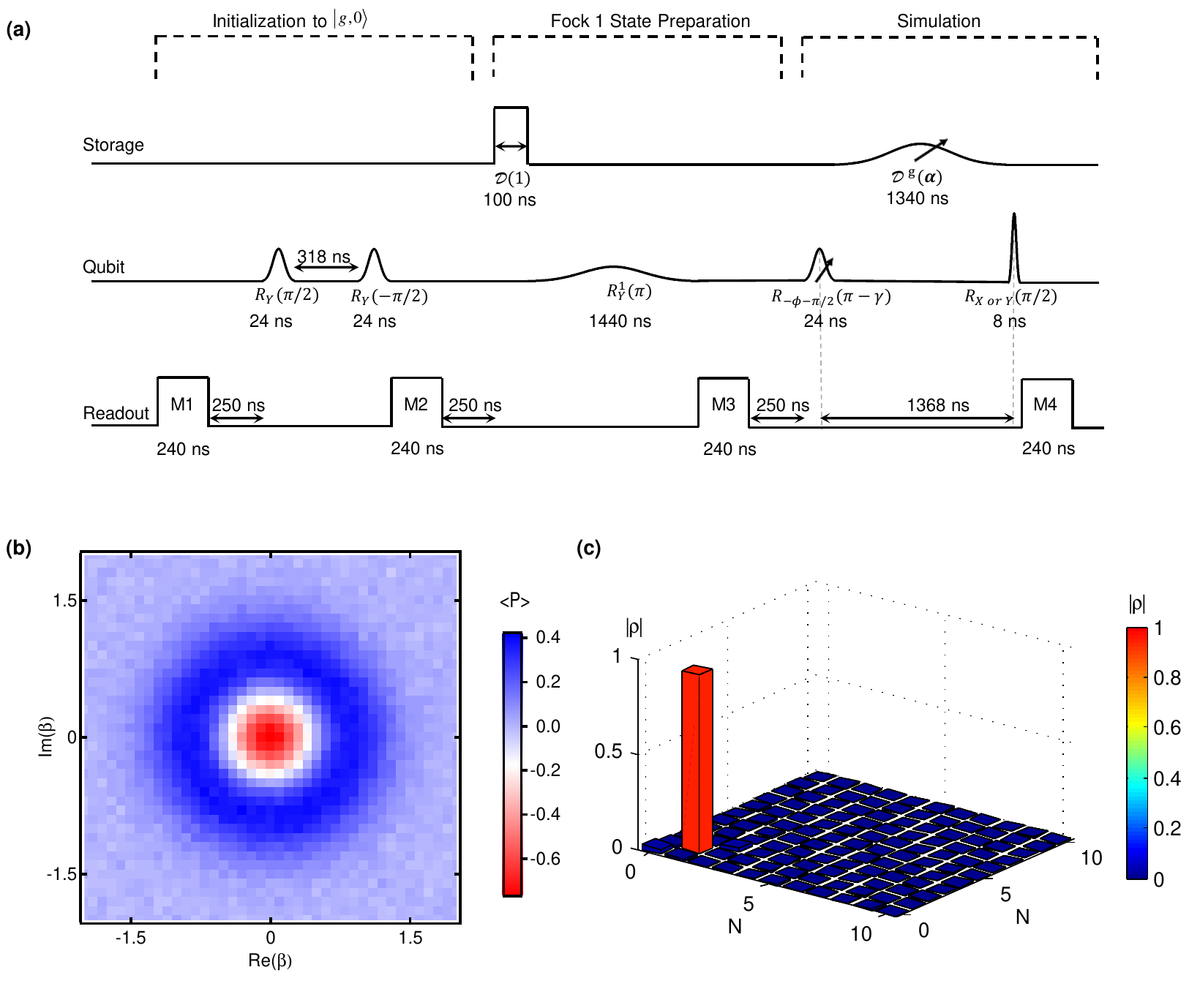}
\caption{\textbf{(a)} Schematic of the pulse sequence for simulating a nuclear initial state $\ket{1}$. The process contains three parts: 1) initialization of the system to $\ket{g,0}$ by post-selecting results of the qubit measurement and the cavity parity measurement; 2) creation of $\ket{1}$ state by first displacing the cavity to $\ket{\alpha=1}$ state, then performing a selective $\pi$ rotation ($\sigma=360$~ns) of the qubit corresponding to $N=1$ photon in the cavity, and finally post-selecting the excited state of the qubit; 3) simulation as described in the main text. In order to minimize the cavity state's Kerr rotation associated with the qubit in the $\ket{e}$ state, the time interval between the last two $\pi/2$ pulses is set to $4\pi/\chi_{qs}$. The last $R_{X or Y}(\pi/2)$ pulse needs as narrow as possible to be non-selective. \textbf{(b)} Wigner function of the created $\ket{1}$ state with a fidelity $F=\left\langle 1\right|\rho\left|1\right\rangle=0.94$. \textbf{(c)} Moduli of the reconstructed density matrix $\rho$ from the Wigner function in \textbf{(b)} by least-square regression using a maximum likelihood estimation. }
\label{fig:Fock1_pre}
\end{figure*}

Figure~\ref{fig:pi_pulse_property} shows the properties of non-selective $\pi$ pulses with a coherent state $\ket{\alpha}$ in the cavity. In our simulation, there is always a coherent state presented in the storage cavity (see Fig.~\ref{fig:Fock1_pre}). The associated measurement pulses are shown in Fig.~\ref{fig:pi_pulse_property}a. The absence or presence of a $\pi$ pulse at the beginning together with  a measurement M1 is used to prepare the qubit's initial state $\ket{g}$ or $\ket{e}$ by a post-selection. The numbers outside the parenthesis correspond to a $\pi$ pulse with $\sigma=6$~ns for $\alpha=1$ ($\bar{n}=1$) in the storage cavity. The numbers inside the parenthesis correspond to a $\pi$ pulse with $\sigma=2$~ns for $\alpha=4$ ($\bar{n}=16$) in the storage cavity. The fidelity loss mainly comes from two parts: the qubit $T_1$ process during the measurement time and the waiting time between the two consecutive measurements, and the qubit frequency shifts due to photon number occupations in the cavity. 
 
{\section{Measurement pulse sequence}}
The pulse sequence for the experiment is shown in Fig.~\ref{fig:Fock1_pre}a. Here we use the case of the nuclear state in a Fock state $\ket{1}$ as an example. The whole experiment can be divided into three main parts: 1) initialization of the system to $\ket{g,0}$ by post-selecting results of the qubit measurement and the cavity parity measurement; 2) creation of $\ket{1}$ state by first displacing the cavity to a coherent state $\ket{\alpha=1}$, then applying a selective $\pi$ rotation ($\sigma=360$~ns) of the qubit corresponding to $N=1$ photon in the cavity, and finally post-selecting the excited state of the qubit; 3) simulation as described in the main text.

The classical microwave pulse to displace the cavity state has either a square envelope with a width of 100~ns or a Gaussian envelope with $\sigma=335$~ns. The amplitudes have been calibrated carefully as shown in the section of ``calibration of driving amplitude". All qubit drive pulses have a Gaussian envelope pulses truncated to $\pm 2\sigma$. To eliminate the possible qubit leakage to higher qubit levels, we also apply the so-called ``derivative removal by adiabatic gate" technique~\cite{Motzoi} for pulses with $\sigma=$2, 4, and 6~ns.

Figure~\ref{fig:Fock1_pre}b shows the Wigner tomography of the created $\ket{1}$ state and Fig.~\ref{fig:Fock1_pre}c shows the moduli of the reconstructed density matrix $\rho$ by least-square regression using a maximum likelihood estimation~\cite{Smolin2012,VlastakisBell}. The measured fidelity of this state is $F=\left\langle 1\right|\rho\left|1\right\rangle=0.94$. The Wigner tomography of the cavity state is performed by a cavity's displacement operation $\cal D(-\beta)$ followed by a parity measurement~\cite{Lutterbach1997,Bertet2002,Vlastakis,SunNature,Liu2016}. The parity measurement is achieved in a Ramsey-type measurement of the qubit with a conditional cavity $\pi$ phase shift $C(\pi)$ sandwiched in between two unconditional qubit rotations $R_{Y}(\pi/2)$ followed by a projective measurement of the qubit.

\section{Calibration of driving amplitude}
\begin{figure*}
\centering
\includegraphics{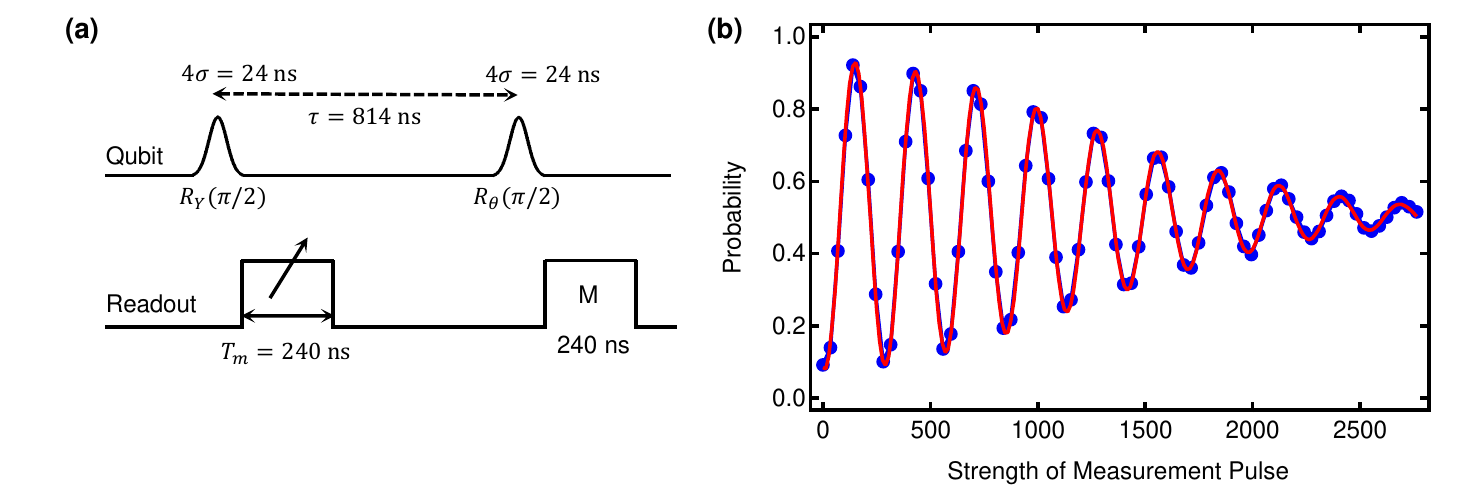}
\caption{Calibration of $\bar{n}$ in the readout pulse. \textbf{(a)} A Ramsey-type measurement sequence for the qubit inserted by a readout pulse with various amplitudes at a fixed width $T_m$. The second $\pi/2$ pulse has a rotating phase relative to the first one in order to have an oscillating interference signal. \textbf{(b)} The probability of the qubit at the ground state as a function of the measurement pulse amplitude. The Ramsey amplitude $A\sim e^{-T_m\Gamma_m}$ gives a calibration of $\bar{n}$ in the readout pulse, where the measurement-induced dephasing rate $\Gamma_m=2\bar{n}\kappa sin^2(tan^{-1}(\chi_{qr}/\kappa))$, $\kappa$ is the readout cavity's decay rate, and $\chi_{qr}$ is the dispersive frequency shift between the readout cavity and the qubit.}
\label{fig:readout_pulse_nbar}
\end{figure*}

To have a high-QND and high-fidelity single-shot readout is essential for any experiment that requires post-selections. There should be enough readout photons for a high signal-to-noise ratio of the readout, but not too many to take a long time for those readout photons to leak out, otherwise slowing down the subsequence operations. In Fig.~\ref{fig:readout_pulse_nbar}, the readout photon number is calibrated through a measurement-induced dephasing process. Figure~\ref{fig:readout_pulse_nbar}a shows the measurement pulse sequence which is a Ramsey-type measurement of the qubit inserted by a readout pulse with various amplitudes at a fixed width $T_m$. Figure~\ref{fig:readout_pulse_nbar}b shows the decaying signal, coming from the dephasing due to readout photons, as a function of the measurement pulse amplitude. An exponential fit gives a calibration of the readout pulse. In our experiment, we typically readout the qubit with about $\bar{n}=5$ photons in the cavity and wait for nearly six times the photon decay time to make sure the average remaining photon number in the readout cavity is only about 1\%.  

\begin{figure*}
\centering
\includegraphics{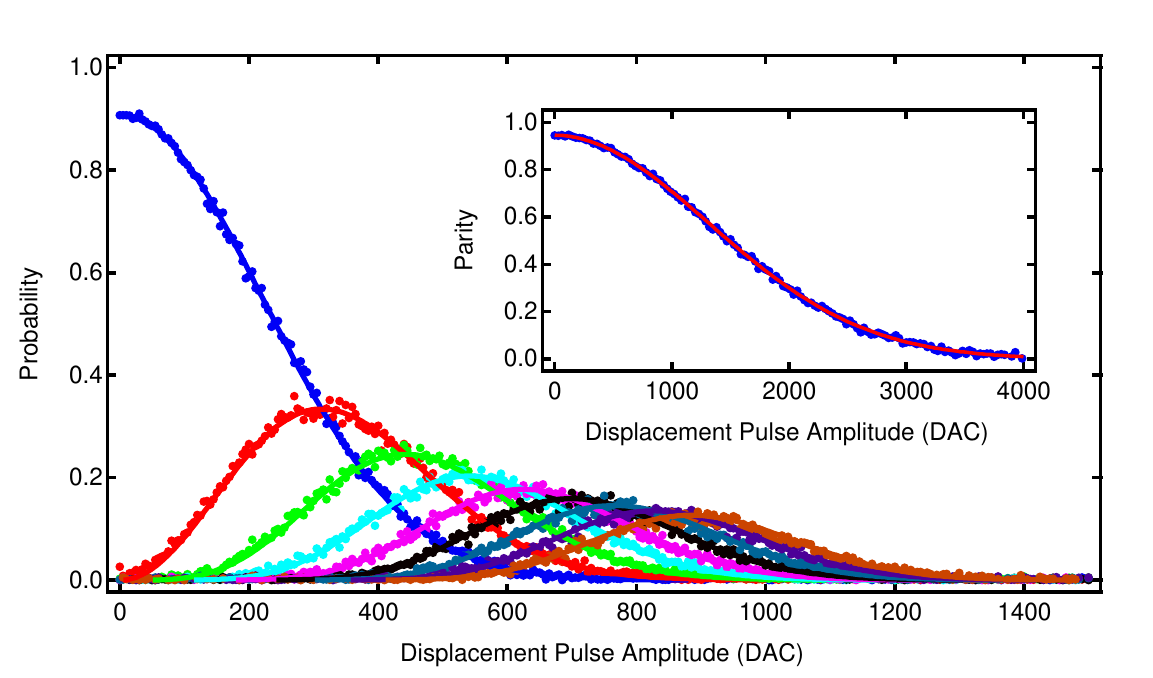}
\caption{Calibration of the controlled cavity displacement ${\cal D}^g(\alpha)$ (Gaussian envelope with $\sigma=335$~ns). The probability of the first nine Fock states $N=0,1,...,8$ is measured as a function of the displacement pulse amplitude through a selective $\pi$ pulse ($\sigma=360$~ns) on the qubit at each resonant transition frequency corresponding to a specific photon number $N$ in the cavity. There is no normalization of the measurement data and the loss of probability dominantly comes from the $T_1$ process in the long duration of the $\pi$ pulse. Lines are from a global fit with a Poisson distribution $\displaystyle P(\ket{\alpha},N)=A|\alpha|^{2N}e^{-|\alpha|^2}/N!$, where $A$ is a scaling factor accounting for the probability loss. The excellent agreement indicates good control of the coherent state in the cavity, giving a calibration DAC level=326 corresponding to $\alpha=1$. Inset: calibration of the displacement ${\cal D}(1)$ with a 100~ns square envelope. This calibration is performed by measuring the photon number parity instead as a function of displacement amplitude, giving DAC level=2649 for $\alpha=1$. The loss of parity measurement contrast mainly comes from the qubit decoherence during the parity measurement.}
\label{fig:calibrate_nbar}
\end{figure*}

The calibration of the controlled cavity displacement ${\cal D}^g(\alpha)$ (Gaussian envelope with $\sigma=335$~ns) is critical for our simulation scheme. This calibration is realized by measuring the probability of the first nine Fock states $N = 0,1,2,...8$ through a selective $\pi$ pulse (Gaussian envelope with $\sigma=360$~ns) as a function of the displacement pulse amplitude, as shown in Fig.~\ref{fig:calibrate_nbar}. A nearly perfect global fit to the measurement results with a Poisson distribution not only gives the required calibration of the cavity displacement amplitude, but also indicates good control of the coherent state in the cavity. The calibration of the displacement ${\cal D}(1)$  with a 100~ns square envelope is performed by measuring the photon number parity instead as a function of displacement pulse amplitude (inset of Fig.~\ref{fig:calibrate_nbar}).

\begin{figure*}
\centering
\includegraphics{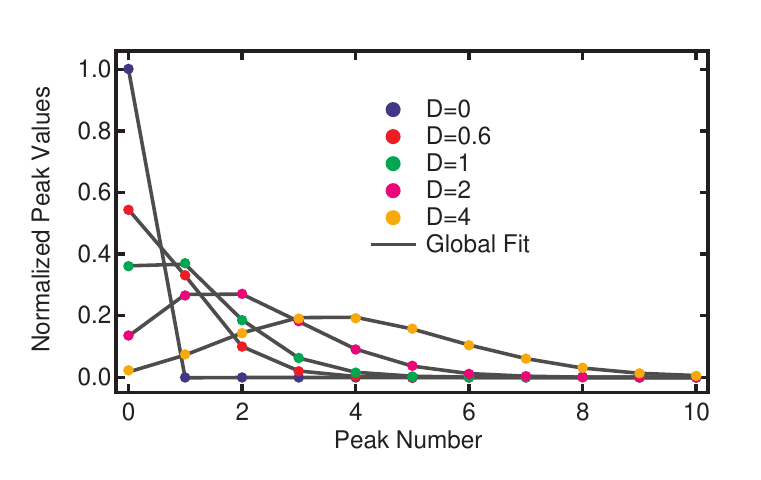}
\caption{Poisson distribution of spectral peaks for the nuclear system at vacuum and at zero temperature. Dots are typical spectral peak values in Fig.~2 of the main text, all normalized by dividing a constant reduction factor $f=0.83$. Lines are from a global fit with a Poisson distribution. The experimental peaks are nearly perfectly Poisson distributed.}
\label{fig:Poisson_Dist}
\end{figure*}

Figure~\ref{fig:Poisson_Dist} shows the typical spectral peak values for the nuclear system at vacuum and at zero temperature (Fig.~2 of the main text). The peaks are all normalized by dividing a constant reduction factor $f=0.83$ (mainly due to the system decoherence), and are in good agreement with the expected Poisson distribution.

